# Mixed-Signal Quantum Circuit Design for Option Pricing Using Design Compiler

*Yu-Ting Kao, Yeong-Jar Chang, Ying-Wei Tseng*
*Industrial Technology Research Institute*

## ABSTRACT

Prior studies [1–5] have largely focused on quantum algorithms, often reducing parallel computing designs to abstract models or overly simplified circuits. This has contributed to the misconception that most applications are feasible only through VLSI circuits and cannot be implemented using quantum circuits. To challenge this view, we present a mixed-signal quantum circuit framework incorporating three novel methods that reduce circuit complexity and improve noise tolerance. In a 12-qubit case study comparing our design with JP Morgan's option pricing circuit [6], we reduced the gate count from 4095 to 392, depth from 2048 to 6, and error rate from 25.86% to 1.64%. Our design combines analog simplicity with digital flexibility and synthesizability, demonstrating that quantum circuits can effectively leverage classical VLSI techniques—such as those enabled by Synopsys Design Compiler—to address current quantum design limitations.

## 1. INTRODUCTION

Quantum computing has demonstrated immense potential due to its unprecedented information capacity and ability to perform massively parallel computations. Pioneering quantum algorithms such as those proposed by Deutsch and Jozsa [1], Simon [2], Shor [3], Grover [4], and Brassard [5] et al. have effectively utilized quantum properties like superposition and entanglement, highlighting significant computational advantages over classical counterparts. However, despite these theoretical advancements, practical quantum circuit implementation still encounters substantial obstacles, including excessive design complexity, susceptibility to noise, and difficulties in accurately translating real-world problems into feasible hardware designs.

Quantum computing can be metaphorically viewed as a "black hole": although it enables large-scale parallel computation, the desired result is often obscured due to the probabilistic nature of measurement. Therefore, quantum circuit design typically involves three key stages: (1) Broadcast: create a broad superposition of quantum states; (2) Calculation: map the problem to a parallel quantum process; and (3) Measurement: Apply specific algorithms (e.g., [1-5]) that reduce the number of quantum states, allowing extraction of the desired outcome with fewer measurement.

Superposition and entanglement pose major challenges in quantum circuit design. While small-scale circuits can be manually created, large-scale systems require automated synthesis tools such as Synopsys Design Compiler (DC). These tools help abstract the complexity of quantum phenomena, enabling more scalable and efficient design. However, in many applications, even after applying DC, the resulting quantum circuits remain too complex to effectively showcase its benefits. Therefore, it is important to identify a clearer example in which DC leads to noticeable improvements.

Both analog circuits [6,7] and quantum neural networks (QNNs) [8,9] face a common limitation: data preprocessing often requires an exponentially growing number of rotation gates. The JP Morgan circuit [6], shown in Figure 1, serves as a compelling case. This paper begins by optimizing that circuit using DC, achieving notable improvements. The case study also shows that mixed-signal circuits, like fully digital ones, can support massively parallel quantum computation.

Figure 1 shows a quantum circuit that computes weighted averages using Quantum Amplitude Estimation (QAE) for option pricing. A classical computer first pre-processes the piecewise linear (PWL) function to isolate the region where the payoff exceeds the strike price. Only this relevant segment is sent to the quantum circuit, which highlights that the quantum computation depends on classically pre-processed input and cannot perform the full payoff calculation independently.

Building on this foundation, we propose three circuit-level improvements: (1) Exponential Data Pre-processing: Aims to reduce the number of quantum gates from $2^n-1$ to $O(n^2)$, and to decrease the computing depth from $2^{n/2}$ to $O(n)$. (2) Digital Calibration: Enhances noise resilience and corrects analog behavior through digital tuning techniques. (3) Monte Carlo Price Simulation: Employs probabilistic models to generate random samples that simulate multi-day asset price fluctuations, followed by the computation of the overall average. The proposed quantum circuit can follow the traditional Monte Carlo method, which uses the sample mean to approximate the population mean, requiring random sampling in the process. Alternatively, it can adopt QAE to achieve quantum advantages, eliminating the need for randomness in the simulation.

The goal of this research is to develop quantum circuits for solving nonlinear financial problems. However, due to ongoing work on nonlinear components—particularly those for exponential and payoff functions—only about 90% of the circuit has been implemented so far. As a result, the current Monte Carlo simulations focus on verifying the linear portions. The averaging operation used is for validation purposes only, not the final objective, ensuring the correctness and efficiency of the proposed circuit. Once the nonlinear components are available, only minor modifications will be required, as most of the proposed circuits can be reused. These linear validations therefore serve as a critical intermediate step toward fully implementing option pricing computations. The nonlinear components themselves are not developed or described in this paper.

By selectively applying DC for circuit optimization, this study demonstrates an effective integration of quantum and classical digital techniques, laying a solid foundation for practical quantum circuit design. Beyond improving quantum circuit efficiency in financial applications, this research also highlights the broader potential of digital optimization tools in enabling scalable and industry-relevant quantum computing solutions.

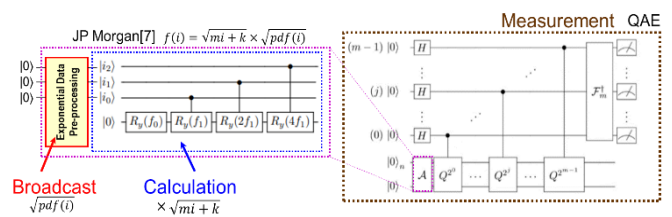

Fig.1 Averaging circuit for option pricing [6]

## 2. METHODOLOGY

### 2.1. Exponential Data Pre-processing

To highlight the bottleneck in analog quantum circuit designs [6–9], we illustrate a simple 3-qubit example, as shown in **Figure 2**. Traditional data preprocessing methods in these designs require an exponentially growing number of rotation gates (up to $2^n-1$) and computational depth (up to $2^{n/2}$). As the number of qubits increases, this exponential growth quickly becomes impractical, severely limiting the scalability of analog quantum circuits. The expected weighted sum used in option pricing can be represented as:

$$\sum_{i=0}^{7}(mi+k)pdf(i) = \sum_{i=0}^{7}(parameter)^2$$

This expression provides the mathematical foundation for the rotation angle encoding and subsequent weighted averaging in the quantum circuit. To overcome these challenges, we applied DC to optimize the circuit architecture. By restructuring the rotation gates, we expect to reduce the overall complexity from exponential to polynomial $O(n^2)$ and the depth to a linear scale $O(n)$, significantly alleviating the bottleneck in data pre-processing.

**Figure 3** further presents the Register Transfer Level (RTL) code corresponding to the encoded lognormal probability distribution. In our implementation, the distribution is discretized into 32 output levels (0 to 31), with inputs ranging from 0 to 4095, representing 12 qubits or quantum states. These RTL-encoded values are mapped to specific thread IDs and serve as inputs to the Design Compiler. The synthesized circuit can perform massively parallel operations across different thread IDs (0 to 4095). Compared to conventional quantum analog circuits, this digitally optimized approach significantly reduces circuit complexity and highlights the practical value of DC-based synthesis in quantum design.

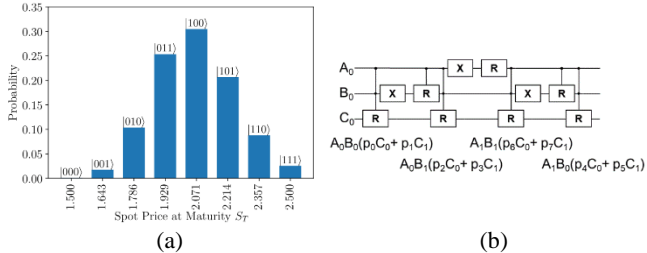

Fig 2. (a) the input probability distribution functions (PDFs) and (b) the corresponding rotation gates

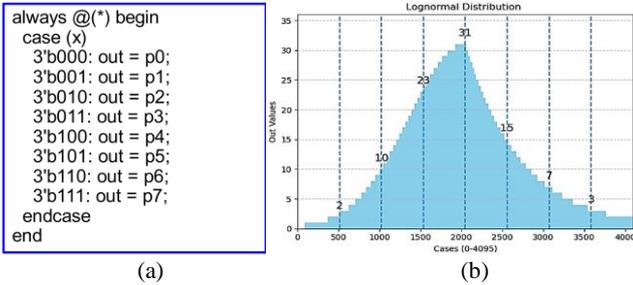

Fig 3. Data pre-processing: (a) RTL code; (b) lognormal distribution

### 2.2. Digital Calibration

This work proposes a novel circuit architecture, called digital calibration, designed to mitigate nonlinear distortions in analog computations. The mechanism can strength out the curve. As illustrated in **Figure 4**, directly feeding the digital input $x_i$ into the analog circuit often results in undesired output signals due to analog inaccuracies. To address this issue, the input $x_i$ is first transformed into a corrected value $r_i$ via a lookup table before being passed to the analog circuit. This preprocessing ensures that the resulting output $y_i$ closely matches the intended value. The proposed method offers a generalized correction framework, with the following digital correction mechanism used for the JP Morgan circuit [6] serving as a specific example.

As shown in **Figure 5**, the rotation gates produce outputs in the form of $sin(x)$, which differ from the desired values of $\sqrt{x}$ required for weighted average calculations. To address this discrepancy, a correction mechanism is introduced to align the output with the expected result. This digital calibration technique computes the appropriate controlling codes by applying the inverse sine function:

$$r = \sin^{-1}(\sqrt{x})$$

With this calibrated angle, the output of the rotation gate becomes:

$$\sin(r) = \sin(\sin^{-1}(\sqrt{x})) = \sqrt{x}$$

JP Morgan's original design constrained rotation gates to small angle variations around 45 degrees, making the circuits particularly sensitive to noise. Small angular intervals such as 0.785, 0.786, 0.787, and 0.788 radians were notably vulnerable to minor disturbances. In contrast, our digitally calibrated approach significantly expands the range of viable rotation angles—for instance, using clearly defined values like 0.1, 0.2, 0.3, and 0.4 radians—thereby greatly enhancing robustness and noise tolerance. This method allows the circuit to reliably tolerate angle deviations of up to approximately 0.05 radians without significant performance degradation.

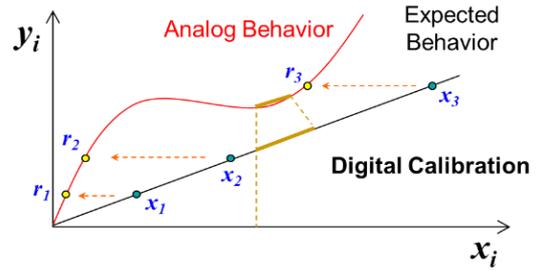

Fig 4. Proposed Digital Calibration Mechanism

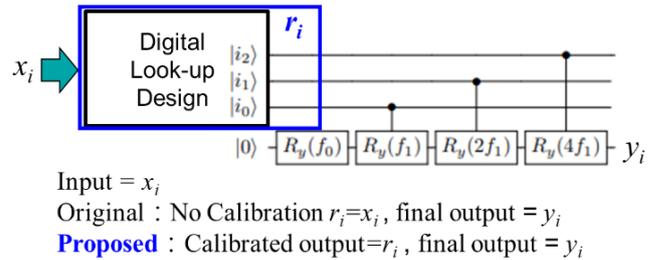

Input = $x_i$
Original : No Calibration $r_i=x_i$, final output = $y_i$
**Proposed** : Calibrated output=$r_i$, final output = $y_i$

Fig 5. Digital lookup table for rounding and adjustment

### 2.3. Monte Carlo Price Simulation

In Monte Carlo-based option pricing, the operations of asset price fluctuations over multiple time steps are typically modeled as multiplications rather than additions. The proposed method simplifies this process by performing a series of additions, such as $a+b+c$, where $a$, $b$, and $c$ may be positive or negative. The exponential operation is deferred to the final stage, where the identity $\exp(a)\times\exp(b)\times\exp(c) = \exp(a+b+c)$ is applied to reconstruct the final asset price. The payoff function is then applied—subtracting the strike price when the final price exceeds it, or assigning zero otherwise. The payoffs from each iteration are finally averaged to compute the expected value.

We present a novel mixed-signal circuit architecture for Monte Carlo-based option pricing. Due to the ongoing development of nonlinear quantum components—specifically those required for exponential and payoff functions—these operations are not yet implemented in this work. Instead, the current design focuses on validating the linear portion of the circuit through averaging techniques. In each iteration, price fluctuations are generated randomly based on a predefined probability distribution function (PDF), accumulated over time, and the final prices are averaged across multiple runs to verify the correctness and practicality of the proposed circuit.

Simulation analysis can prove that the newly invented quantum circuit is consistent with the traditional Monte Carlo results. This demonstrates that our circuit achieves an averaging capability comparable to that of J.P. Morgan's approach. Moreover, compared to fully digital quantum designs, the proposed circuit is more streamlined and can be further optimized using commercial electronic design automation tools such as DC.

The proposed circuit architecture includes a quantum random number generator to simulate daily fluctuations in the underlying asset price. These fluctuations are then accumulated using a single-qubit price accumulator composed of rotation gates (**Fig. 6**). This circuit enables the quantum circuit to capture cumulative changes.

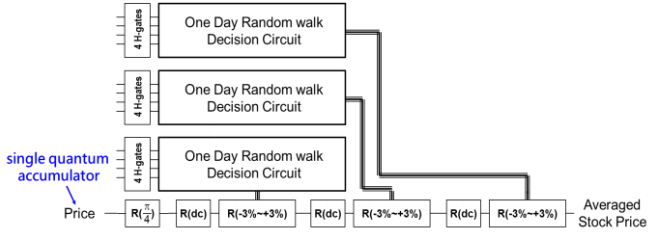

Fig 6. Mixed-signal circuit for Monte Carlo price simulation

In contrast to JP Morgan's method, which evaluates the function $f(i) = \sqrt{(mi+k)} \times \sqrt{(pdf(i))}$ for all indices $i = 0, 1, 2, \ldots, 2^n-1$, our approach achieves massive parallelism by simultaneously executing multiple simulation iterations, thereby aligning more closely with the principles of classical Monte Carlo methods. Furthermore, this approach eliminates the need to explicitly construct the probability distribution function $pdf(i)$ described above, thereby reducing statistic complexity and avoiding potential estimation errors. The proposed circuit supports two measurement strategies: the first yields the same results as classical Monte Carlo through direct measurement, while the second applies QAE with fewer measurements to achieve quantum advantages.

Ideally, the quantum operation should reflect the value of $\sqrt{x}$ to facilitate accurate averaging of stock prices. However, the rotation gates in the circuit inherently produce outputs in the form of $\sin(y)$. Therefore, a consistent linear transformation is required to map $\sin(y)$ to $\sqrt{x}$ across varying rotation angles. This requirement mirrors the behavior observed in the prior work by JP Morgan [6], where the approximation holds only under the assumption of small-angle variations centered around a 45-degree rotation.

To ensure accuracy across all thread IDs with varying rotation angles, an initial assumption employed a simple linear scaling model ($x=\pi/4+m\theta$, $y=\pi/4+\theta$), where the scaling factor m=2 was heuristically set based on square-root and double-angle trigonometric identities. However, subsequent analysis demonstrated that this approximation led to larger errors over a broader range of rotation angles, highlighting the need for a more accurate approach—referred to as **analog calibration**. Through both empirical evaluation and theoretical validation, we identified m=1.57 as the optimal scaling factor that minimizes the approximation error, as shown below:

$$f(\theta) = \frac{\sqrt{\frac{\pi}{4} + m\theta}}{\sin(\frac{\pi}{4} + \theta)} = \frac{\sqrt{\frac{\pi}{4}} \times \sqrt{1 + \frac{4m\theta}{\pi}}}{\sin\frac{\pi}{4}\cos\theta + \cos\frac{\pi}{4}\sin\theta}$$

$$\approx \frac{\frac{\sqrt{\pi}}{2}\left(1 + \frac{2m\theta}{\pi}\right)}{\frac{\sqrt{2}}{2} + \frac{\sqrt{2}}{2}\theta} = \sqrt{\frac{\pi}{2}} \times \frac{1 + \frac{2m\theta}{\pi}}{1 + \theta} = \sqrt{\frac{\pi}{2}}$$

At the final stage of the derivation, setting $m=\pi/2 \approx 1.57$ causes $f(\theta)$ to approach a constant, indicating that the error from the linear transformation is nearly zero.

Traditional random number generation relies on hardware (e.g., linear feedback shift registers) or software algorithms to ensure uniform randomness. In contrast, quantum circuits are inherently deterministic and do not produce randomness by default. The proposed quantum circuit (Figure 6) introduces randomness through Hadamard gates followed by direct measurements. Each measurement projects the quantum state onto a probabilistic basis, effectively producing random outputs. The circuit performs m Hadamard operations and measurements, simulating classical Monte Carlo m independent random samples. This demonstrates how quantum circuits can generate randomness for parallel operations.

When QAE is used, the simulation becomes fully deterministic by averaging over all quantum states, removing the randomness effect. For instance, with 20 major qubits and one million ($2^{20}$) parallel runs, each run may involve generating a random value $r = 0, 1, 2, \ldots 15$ using four Hadamard gates. However, QAE evaluates all $2^{24}$ possible states simultaneously, making the randomness unobservable.

Simulation results confirm that the proposed design behaves similarly to classical Monte Carlo methods, producing accurate average estimations. **Figure 7** illustrates an implementation for modeling risk-free interest rates and price fluctuations.

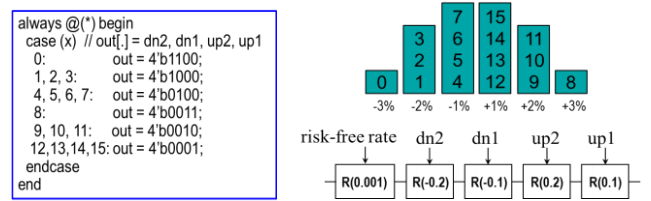

Fig 7. Circuit implementation for risk-free interest rate and price fluctuations

This work extends the use of quantum circuits to option pricing and contributes to mixed-signal circuit design. It shows that massively parallel computation is possible not only in digital quantum systems but also in mixed-signal ones. In our method, the multi-bit digital price accumulator is replaced by a single-qubit version, while the rest of the architecture stays the same as in digital quantum Monte Carlo. These results show that mixed-signal quantum circuits are practical and efficient for real-world financial applications.

### 2.4. The Proposed Circuit for Option Pricing

Building upon the foundational concepts presented in Sections 2.1 and 2.2, we propose the quantum circuit architecture depicted in **Figure 8**. In this design, exponential data pre-processing is implemented by DC to achieve a notable reduction in circuit complexity. Additionally, two digital calibration modules, Arcsin, are integrated to ensure accurate rotation operations and to enhance noise immunity across a wide range of operational angles. These improvements are benchmarked against the circuit architecture proposed by JP Morgan [6], demonstrating superior performance in terms of both precision and hardware efficiency. Although the multiplication of probability density functions (pdf) with linear inputs (lin) can be implemented using digital techniques, the resulting design would be prohibitively complex. Instead, we adopt the methodology proposed by Vazquez[10], significantly simplifying implementation and reducing overall design complexity.

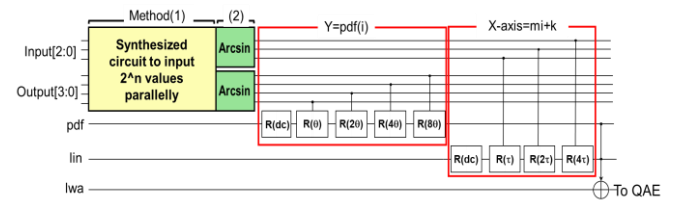

Fig 8. The proposed circuit for financial pricing using data pre-processing and digital calibration

**Figure 9** is the proposed circuit for Monte Carlo price simulation with the ability to simulate daily stock price fluctuations based on a predefined probabilistic model, randomly emulate upward or downward price movements. Instead of employing multi-bit digital accumulators to track the cumulative price changes, the design utilizes a single-qubit price accumulator. This approach not only simplifies the circuit architecture but also allows direct interfacing with the QAE module for final measurement. The resulting output corresponds to the average simulated stock price, thereby achieving the same computational functionality as the quantum circuit proposed by JP Morgan.

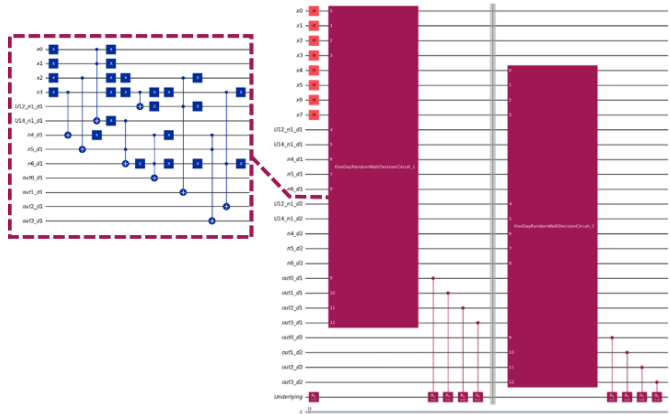

Fig 9. Quantum circuit for Monte Carlo price simulation

## 3. EXPERIMENTAL RESULTS

The Exponential Data Pre-processing circuit significantly reduces gate complexity, lowering the quantum gate count from 4095 to 392 and the depth from 2048 to 6, effectively resolving the challenge of inputting massive amounts of data into quantum circuits.

Digital Calibration usage range of the Rotation Gate is expanded between 0 to $\pi/2$ radians. Through Digital Calibration, the maximum calculation error was notably reduced from 25.86% to merely 1.64% **(Fig. 10)**, greatly improving circuit reliability and robustness against noise.

The Monte Carlo Price Simulation implemented via Mixed-Signal Design replaces the conventional digital multi-bit accumulator with a simpler single qubit accumulator. This design allows direct QAE measurements, significantly reducing overall design complexity. The simulation outcomes closely match those obtained from traditional Monte Carlo analyses, verifying the accuracy and practical feasibility of the proposed quantum approach.

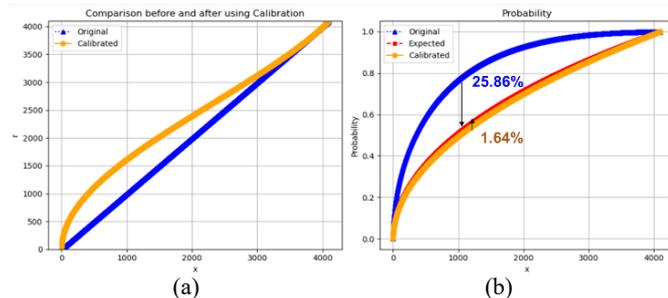

Fig 10. Calibration comparison: (a) before and after; (b) maximum calculation error

Consistency is demonstrated with classical Monte Carlo analysis. Specifically setting a zero probability for unchanged asset prices (+0%) to validate quantum and traditional computation consistency **(Fig. 11)**. Although the histogram analysis is performed using Qiskit, Figures 11 and 12 are generated from the Statevector, not from random measurement Shots. For example, the statistical parameter k=0.795 (representing a stock price on the x-axis) appears 4 times, while k=0.805 appears 3 times, and so on. This histogram serves as a visual representation to facilitate comparison with classical computational results. However, it is important to note that this histogram analysis is not an intrinsic part of the quantum computation process. Within the quantum framework, QAE directly computes the average stock price by aggregating the results of parallel simulations, without relying on histogram-based statistical processing. In the current design, only linear computations are performed to obtain the average; future work will focus on extending the architecture to support nonlinear computations.

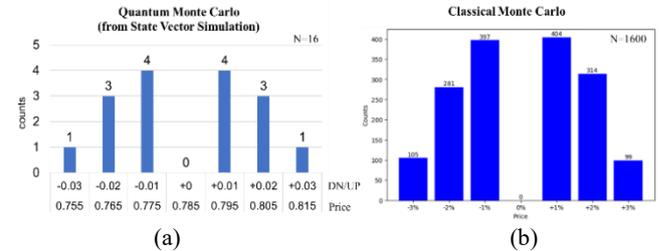

Fig 11. Random Walk Generator (1 day) Price distribution: (a) Quantum Monte Carlo, (b) Classical Monte Carlo

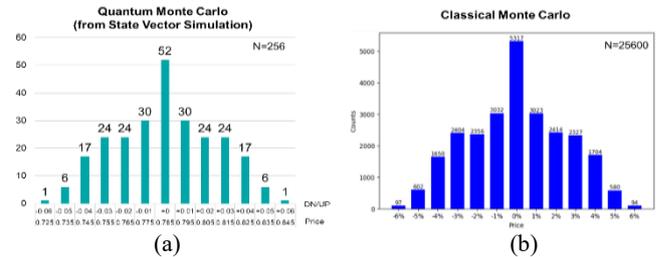

Fig 12. Random Walk Generator (2 day) Price distribution: (a) Quantum Monte Carlo, (b) Classical Monte Carlo

## 4. CONCLUSION

This paper presents novel mixed-signal quantum circuits for Monte Carlo option pricing, combining the simplicity of analog design with the flexibility and synthesizability of digital components. The proposed circuits incorporate three key techniques: Exponential Data Pre-processing, Digital Calibration, and Monte Carlo Price Simulation.

**Exponential Data Pre-processing** reduces circuit complexity when encoding large datasets, decreasing the number of quantum gates from 4095 to 392 and computational depth from 2048 to 6, for a 12-qubit input.

**Digital Calibration** extends the operable range of quantum rotation gates and significantly improves noise tolerance, reducing the maximum error rate from 25.86% to 1.64% through accurate LUT-based calibration.

**Monte Carlo Price Simulation** leverages quantum randomness and a single-qubit price accumulator to model multi-day asset price movements, producing results consistent with classical simulations while benefiting from quantum parallelism.

Although the current implementation does not yet support non-linear functions such as exponential and payoff calculations, the architecture is designed for future extensibility. If these functions must remain digital, only the Single-Qubit Accumulator needs to be replaced. If analog implementations become feasible, the entire design can be preserved. In either case, the approach remains practically valuable. Using JP Morgan's quantum option pricing circuit [6] as a baseline, this work demonstrates how digital synthesis techniques can significantly simplify large-scale quantum circuit design, thereby highlighting the effectiveness of the Synopsys Design Compiler (DC). The proposed architecture contributes to quantum computing in finance and other real-world applications.